%% file: conf_rhopi_eprint.tex
\documentclass[11pt]{article}
\usepackage{graphicx}
\usepackage{epsf}
\usepackage{rotating}
\usepackage{epsfig}
\usepackage{supertabular}
\usepackage{subfigure}

\newcommand{\BABARPubYear}    {02}

\newcommand{\BABARConfNumber} {033}
\newcommand{\SLACPubNumber} {9303}

\input pubboard/babarsym
\input{Definitions}

\long\def\inst#1{\par\nobreak\kern 4pt\nobreak
    {\it #1}\par\vskip 10pt plus 3pt minus 3pt}

\begin{document}
{\pagestyle{empty}

\begin{flushright}
\babar-CONF-\BABARPubYear/\BABARConfNumber \\
SLAC-PUB-\SLACPubNumber \\
July 2002 \\
\end{flushright}

\par\vskip 3cm

\begin{center}
\Large \bf \boldmath Search for {\em CP} Violation in \Bz/\Bzb Decays to $\pi^{+}\pi^{-}\pi^0$ and $K^{\pm}\pi^{\mp}\pi^0$ \\
in Regions Dominated
by the $\rho^{\pm}$~Resonance
\end{center}
\bigskip

\begin{center}
\large The \babar\ Collaboration\\
\mbox{ }\\
July 24, 2002
\end{center}
\bigskip \bigskip

\begin{center}
\large \bf Abstract
\end{center}

\noindent
We present preliminary measurements of direct and indirect \CP-violating 
asymmetries in the decays of neutral $B$ mesons to $\pi^{+}\pi^{-}\pi^0$ 
and $K^{\pm}\pi^{\mp}\pi^0$ final states
dominated by the $\rho^{\pm}$~resonance, 
using a time-dependent maximum likelihood analysis. 
The data sample comprises $88 \times 10^6\,\, \FourS \to B\Bbar$
decays ($80.8\,\fb^{-1}$) collected with 
the \babar\ detector at the \pep2 asymmetric-energy $B$~Factory at SLAC.
For the \CP\ violation parameters, we measure
\begin{eqnarray*}
 \AcprhoK = 0.19\,\pm 0.14\,\,\,{\rm(stat)}\,\,\, \pm{0.11}\,\,\, {\rm(syst)},
& & 
\Acprhopi = -0.22\,\,\,\pm 0.08\,\,\,{\rm(stat)}\,\,\, \pm{0.07}\,\,\, {\rm(syst)}, \\ 
 C_{\rho\pi} \,\,\, = \,\,\,0.45\,\,\,^{+0.18}_{-0.19}\,\,\,{\rm(stat)}\,\,\, \pm{0.09}\,\,\, {\rm(syst)}, 
& & 
 S_{\rho\pi}\,\, =\,\,\,\,\, 0.16\,\,\,\pm 0.25\,\,\,{\rm(stat)}\,\,\, \pm{0.07}\,\,\, {\rm(syst)}.
\end{eqnarray*}
For the other parameters in the description of the $\Bz(\Bzb) \to \rhopi$
decay-time dependence, we obtain
\begin{eqnarray*}
\dC_{\rho\pi} \, = \, 0.38\,\,\,^{+0.19}_{-0.20}\,\,\,{\rm(stat)}\,\,\, \pm{0.11}\,\,\, {\rm(syst)},
&&
\dS_{\rho\pi} =0.15\,\,\,\pm 0.26\,\,\,{\rm(stat)}\,\,\, \pm{0.05}\,\,\, {\rm(syst)}. \\
\end{eqnarray*}

\vfill
\begin{center}
Contributed to the 31$^{st}$ International Conference on High Energy Physics,\\ 
7/24---7/31/2002, Amsterdam, The Netherlands
\end{center}

\vspace{1.0cm}
\begin{center}
{\em Stanford Linear Accelerator Center, Stanford University, 
Stanford, CA 94309} \\ \vspace{0.1cm}\hrule\vspace{0.1cm}
Work supported in part by Department of Energy contract DE-AC03-76SF00515.
\end{center}

\newpage
} 

\input pubboard/authors_ICHEP2002.tex


\input{Introduction}

\input{Detector}

\input{Selection}

\input{BBackground}

\input{ML}

\input{Results}

\input{Systematics}

\input{Summary}

\section{Acknowledgments}
\label{sec:Acknowledgments}


\input pubboard/acknowledgements

\input{Biblio}

\end{document}

%% file: Definitions.tex
\newcommand{\bit}{\begin{itemize}}
\newcommand{\eit}{\end{itemize}}
\newcommand{\beq}{\begin{equation}}
\newcommand{\eeq}{\end{equation}}
\newcommand{\beqn}{\begin{eqnarray}}
\newcommand{\eeqn}{\end{eqnarray}}
\newcommand{\beqns}{\begin{eqnarray*}}
\newcommand{\eeqns}{\end{eqnarray*}}

\def\rar{\rightarrow}

\def\btorhc{\Bz \rar \rho^\pm h^\mp}

\def\rhopi{\rho\pi}
\def\rhoX{\rho h}
\def\rhok{\rho K}

\def\AcprhoX{A_{CP}^{\rhoX}}
\def\Acprhopi{A_{CP}^{\rhopi}}
\def\AcprhoK{A_{CP}^{\rho K}}

\def\hel{{\theta_{\pi}}}
\def\cshelrho{{\rm cos}\,\hel}

\def\de{\Delta E}
\def\mes{\ensuremath{m_{\rm ES}}}

\def\MeVc2{${\rm MeV}/c^2$}
\def\GeVc2{${\rm GeV}/c^2$}

\def\babar{$\mbox{\sl B\hspace{-0.4em} {\small\sl A}\hspace{-0.37em} 
\sl B\hspace{-0.4em} {\small\sl A\hspace{-0.02em}R}}$}

\def\BB{$B\overline{B}$}

\def\BR{{\cal BR}}

\def\NN{{\rm NN}}

\def\diffD {\ensuremath{\Delta D}}

\def\fBzxpm {\ensuremath{f^{\rho^\pm h^\mp}_{\Bz {\rm tag}}}}
\def\fBzbxpm {\ensuremath{f^{\rho^\pm h^\mp}_{\Bzb {\rm tag}}}}

\def\Deltat {\ensuremath{\Delta t}\xspace}
\def\mes {\ensuremath{m_{ES}}\xspace}

\def\Dbar{\ensuremath{\overline{D}}\xspace}
\def\S{\ensuremath{S}\xspace}

\def\dm {\ensuremath{\Delta m_d}\xspace}
\def\dt {\ensuremath{\Delta t}}

\def\dC{\ensuremath{\Delta C}\xspace}
\def\dS{\ensuremath{\Delta S}\xspace}

%% file: pubboard/authors_ICHEP2002.tex
\begin{center}
\small

The \babar\ Collaboration,
\bigskip

B.~Aubert,
D.~Boutigny,
J.-M.~Gaillard,
A.~Hicheur,
Y.~Karyotakis,
J.~P.~Lees,
P.~Robbe,
V.~Tisserand,
A.~Zghiche
\inst{Laboratoire de Physique des Particules, F-74941 Annecy-le-Vieux, France }
A.~Palano,
A.~Pompili
\inst{Universit\`a di Bari, Dipartimento di Fisica and INFN, I-70126 Bari, Italy }
% ...

\end{center}\newpage

%% file: Introduction.tex
\section{Introduction}
\label{sec:Introduction}
In the Standard Model, $\CP$-violating effects arise from a single
complex phase in the three-generation CKM quark-mixing matrix~\cite{CKM}.  
One of the central questions in particle physics is whether this
mechanism is sufficient to explain the pattern of \CP violation observed 
in nature. Recent measurements of the parameter $\stwob$ by the
\babar\/~\cite{bib:BabarS2b,bib:BabarS2b2002} and Belle~\cite{bib:BelleSin2betaObs,bib:BelleEtapKs} 
Collaborations establish that \CP\ symmetry is violated in the neutral $B$-meson 
system. In addition, these two experiments have studied \CP-violating asymmetries 
in $B$ decays to the charmless two-body final states 
$\Kp\pim$ and $\pip\pim$~\cite{bib:BabarSin2alpha,bib:BelleSin2alpha}. 
The time-dependent asymmetry in $\pip\pim$ is related to the angle $\alpha$ of the
unitarity triangle.

In this paper, we investigate \CP\ violation using charmless \Bz/\Bzb~decays to
$\pi^{+}\pi^{-}\pi^0$ and $K^{\mp}\pi^{\pm}\pi^0$ dominated by the
$\rho^{\pm}h^{\mp}$ intermediate state, where $h=\pi$ or $K$. As
in the case of $\pip\pim$, the $\rhopi$ mode provides a probe of both
direct \CP\ violation and \CP\ violation in the interference between
mixing and decay amplitudes.  The latter type of \CP\ violation is
related to the angle $\alpha$. In contrast to $\pip\pim$,
$\rho^{\pm}\pi^{\mp}$ is not a \CP\ eigenstate and four configurations
$(\Bz(\Bzb) \to \rho^{\pm}\pi^{\mp})$ have to be considered.  Although
this leads to a more complex analysis~\cite{bib:thNonCpEig}, it
benefits from a higher branching fraction $(20-30 \times
10^{-6})$~\cite{bib:BaBarRhopi, bib:BelleRhopi}.

The $\rho$ resonance is broad ($150~\mevcc$)  and 
the  $\rho^{\pm}\pi^{\mp}$ state 
may receive contributions at the amplitude level
from other decay channels ({\em e.g.}, $B^0\rightarrow \rho^{\prime +}\pi^-$).
For this analysis, we restrict ourselves to the two regions of the $h^{\pm}\pi^{\mp}\pi^0$ 
Dalitz plot dominated by 
$\rho h$ and assign a label, $\rho^{+}h^{-}$ or 
$\rho^{-}h^{+}$,
to each event depending on the kinematics of the  
$h^{\pm}\pi^{\mp}\pi^0$ final state.
In the following, we will use the $\rho^{+}h^{-}$ 
or $\rho^{-}h^{+}$ labels 
with the above meaning.

Defining $\deltat = t_{\rhoX} - t_{\rm tag}$ as the time interval 
between the decay of $B^0_{\rhoX}$ and that of the other $B^0$ meson in the event, 
$B^0_{\rm tag}$, the decay rate distributions can be written as~\cite{bib:BaBarPhysBook}
\begin{eqnarray}
\label{eq:thTime}
	\fBzxpm(\deltat)\!\!\!\!\! 
	& =& \!\!\!\!\! (1\pm \AcprhoX)
	\frac{e^{-\left|\deltat\right|/\tau}}{4\tau} 
		\bigg[1 + \bigg(
		  (S_{\rho h}\pm\dS_{\rho h})\sin(\deltamd\deltat)
		  -(C_{\rho h}\pm\dC_{\rho h})\cos(\deltamd\deltat)
	\bigg)
		\bigg],\nonumber\\
	\fBzbxpm(\deltat)\!\!\!\!\! 
	& =& \!\!\!\!\! (1\pm \AcprhoX)
	\frac{e^{-\left|\deltat\right|/\tau}}{4\tau} 
		\bigg[1 - \bigg((S_{\rho h}\pm\dS_{\rho h})\sin(\deltamd\deltat)
		  -(C_{\rho h}\pm\dC_{\rho h})\cos(\deltamd\deltat)
	\bigg)
		\bigg].
		\nonumber\\
\end{eqnarray}

The time-integrated charge asymmetries $\Acprhopi$ and $\AcprhoK$
measure direct \CP\ violation.  The time dependence is described by
four additional parameters. In the case of the self-tagging $\rhok$
mode, the values of these four parameters are known to be $C_{\rho K}
= 0$, $\dC_{\rho K} = -1$, $S_{\rho K} = 0$, and $\dS_{\rho K} = 0$.
For the $\rhopi$ mode, they allow us to probe \CP\ violation.  Summing
over the $\rho$ charge in Eq.~\ref{eq:thTime}, and neglecting the
charge asymmetry~$A_{CP}^{\rho\pi}$, one obtains the simplified \CP\ asymmetry between
the number of $\Bz$ and $\Bzb$ tags, given by
\begin{equation}
A_{\Bz/\Bzb} = (N_{B^0} - N_{\bar{B^0}})/(N_{B^0} + N_{\bar{B^0}}) 
\sim S_{\rho\pi}\sin(\deltamd\deltat) -C_{\rho\pi}\cos(\deltamd\deltat) \; .
\label{eq:Assym}
\end{equation}
The parameter $C_{\rho\pi}$ describes the time-dependent direct \CP violation and $S_{\rho\pi}$
 measures \CP\ violation in the interference between mixing and decay 
related to the angle $\alpha$.

The parameters $\dC_{\rho \pi}$ and $\dS_{\rho\pi}$ are insensitive to
\CP violation. The asymmetry between $N({\Bz_{\rhopi}} \to
{\rho^+\pi^-}) + N({\Bzb_{\rhopi}} \to {\rho^-\pi^+})$ and
${N(\Bz_{\rhopi}} \to {\rho^-\pi^+}) + N({\Bzb_{\rhopi}} \to
{\rho^+\pi^-})$ is described by $\dC_{\rho \pi}$, while
$\dS_{\rho\pi}$ is sensitive to the strong phase difference between
the amplitudes contributing to $\Bz \to\rhopi$ decays. The naive
factorization model~\cite{bib:thNonCpEig} predicts $\dC_{\rho
\pi}\sim0.4$ while there is no prediction for $\dS_{\rho\pi}$.


The measurements of the six parameters $\AcprhoK$, $\Acprhopi$, $C_{\rho\pi}$, $\dC_{\rho\pi}$,
$S_{\rho\pi}$, and $\dS_{\rho\pi}$ reported here are
 performed using events collected by the 
\babar\ detector at the \pep2 asymmetric-energy $B$~Factory between January 2000 and June 2002. 
This sample corresponds to
an integrated luminosity of 80.8~\invfb taken 
at the \FourS resonance (``on-resonance''), which represents 88 million $B\Bbar$ pairs,
and 9.6~\invfb taken around 40 \mev below the resonance (``off-resonance'').

We extract the yields and $\CP$~parameters using a time-dependent maximum likelihood analysis
based on Eq.~\ref{eq:thTime}. 
This paper is organized as follows: the \babar\ detector is described briefly in 
Sec.~\ref{sec:detector}. The event reconstruction and selection procedure 
is given in Sec.~\ref{sec:Selection}. 
$B$-related backgrounds and their treatment in the likelihood analysis 
are described in Sec.~\ref{sec:BBackground}. 
The full maximum likelihood fit is discussed in Sec.~\ref{sec:ML}. Finally, 
the results and the evaluation 
of systematic uncertainties are given in Secs.~\ref{sec:Results} and~\ref{sec:Systematics}.

%% file: Detector.tex
\section{The \babar~Detector}
\label{sec:detector}

A detailed description of the \babar~detector can be found in
Ref.~\cite{bib:babarNim}.  Charged particle momenta are measured in a
tracking system consisting of a $5$-layer double-sided silicon vertex
tracker (SVT) and a $40$-layer drift chamber (DCH) filled with a gas
mixture based on helium and isobutane. The SVT and DCH operate within
a $1.5$-T superconducting solenoidal magnet. The typical decay vertex
resolution is around $65~\mu {\rm m}$ along the beam direction for the
fully reconstructed~$B^0_{\rhoX}$, and around $100$ to $150~ \mu {\rm
m}$ for the partially reconstructed tagging $B^0_{\rm tag}$.  Photons
are detected in an electromagnetic calorimeter (EMC) consisting of
$6580$ CsI(Tl) crystals arranged in barrel and forward end-cap
sub-detectors.  The $\pi^0$ mass resolution
is on average $7 \mevcc$.  The flux return for the
solenoid is composed of multiple layers of iron and resistive plate
chambers for the identification of muons and long-lived neutral
hadrons.  Tracks from the signal $B$ decay are identified as pions or
kaons by the Cherenkov angle $\theta_{Ch}$ measured with a detector of
internally reflected Cherenkov light (DIRC). The typical separation
between pions and kaons varies from $8\sigma$ at $2\,\gevc$ to
$2.5\sigma$ at $4\,\gevc$, where $\sigma$ is the average $\theta_{Ch}$
resolution. Lower momentum kaons are identified with a combination of
$\theta_{Ch}$ (for momenta down to $0.7\,\gevc$) and measurements of
ionization energy loss, $dE/dx$, in the DCH and SVT.

%% file: Selection.tex

\section{The Event Selection and Reconstruction}
\label{sec:Selection}
\label{sec:tagging}

Signal $B_{\rho h}$ candidates are reconstructed from combinations 
of two charged tracks and a $\pi^0$~candidate.
The charged tracks are required to be inconsistent with being an electron based on $dE/dx$
measurements,  shower shape criteria
in the EMC, and the ratio of shower energy and track momentum. The photons 
from the $\pi^0$ must have an energy greater than $50\mev$, and a lateral 
shower profile variable~\cite{bib:LAT} between $0.01$ and $0.6$.
 The invariant mass $m(\gamma\gamma)$ of the photons 
must satisfy $0.11<m(\gamma\gamma)<0.16~\gevcc$. 
Similarly, to form a $\rho$ candidate, the invariant mass $m(\pi^\pm\pi^0)$ of the
charged track and $\pi^0$  must satisfy $0.4<m(\pi^\pm\pi^0)<1.3~\gevcc$. 
If both the $(\pi^+\pi^0)$ and $(\pi^-\pi^0)$ pairs satisfy this requirement, 
the $B$~candidate is rejected, as the
$\pi^+\pi^-\pi^0$ might result from interfering $\rho$'s, and cannot be associated 
with a definite $\rho$ charge. 
The track used for the $\rho$ candidate must be inconsistent with being 
a kaon based on $dE/dx$ and DIRC information. Finally, we require  
$|\cshelrho| > 0.25$, where
$\hel$ is the angle between the charged pion in the rest frame of the $\rho$ 
and the $\rho$ flight direction 
in the rest frame of the $B$. We refer to the track~$h$ in~$\rhoX$ as the {\em bachelor track}. 
To reject two-body $B$-background, 
the invariant mass of the two charged tracks, and the invariant mass of the bachelor track 
and the $\pi^0$ must be less than $5.14~\gevcc$.

Two kinematic variables, used in the maximum likelihood fit, allow 
discrimination of signal $B$~candidates from fake $B$ candidates 
due to random combinations of tracks 
and $\pi^0$~candidates. 
The first variable is the beam-energy substituted mass defined as 
\begin{equation}
\mes = \sqrt{(s/2 + {\mathbf {p}}_i\cdot {\mathbf {p}}_B)^2/E_i^2 - {\mathbf {p}}_B^2},
\end{equation}
where $s$ is the square of the center-of-mass (CM) energy, $E_i$ and ${\mathbf {p}}_i$
are the total energy and three-momentum of the \epem\ state in the
laboratory frame, and ${\mathbf {p}}_B$ is the three-momentum of the
$B$ candidate in the same frame. Signal events populate the $\mes$
region around the $B$ mass  with a peak resolution of
around $2.6\mevcc$. Candidates are required to satisfy $5.23 < \mes <
5.29~\gevcc$. The second variable, $\de=E_B^*-\sqrt{s}/2$, is the difference between
the reconstructed energy of the $B_{\rho h}$ candidate in the CM frame
and the beam energy. The $\de$ distribution
for signal events with a pion bachelor track ($\rhopi$) peaks around zero, while
the distribution for $\rhok$~signal events with the $\pi$ mass hypothesis assigned to the
true kaon track, is shifted by $-45$ MeV on average (the exact shift
depends on the momentum of the kaon). Backgrounds from other decay
modes of the~$B$ peak at different $\de$ depending on the number of
charged and neutral particles in the decay: two-body decays, three-body decays,
and four-body decays peak at positive, approximately~$0$, and
negative~$\de$, respectively. In order to reduce background from other
modes, we require $-0.12<\de<0.15~\gev$.

The time difference $\dt$ is obtained from the measured distance between the $z$ positions (along the beam direction) 
of the $\Bz_{\rhoX}$ and $\Bz_{\rm tag}$ decay vertices, and the known boost of the \epem\ system. The vertex of the 
$\Bz_{\rm tag}$ is reconstructed from all tracks in the event except those from the $\Bz_{\rhoX}$, and an iterative
procedure~\cite{bib:BabarS2b} is used to remove tracks with a large contribution to the vertex~$\chi^2$. 
An additional constraint is obtained 
from the three-momentum and vertex position of the $\Bz_{\rhoX}$ candidate, and the average \epem\ interaction point 
and boost. We require $|\dt| < 20~{\rm ps}$ and $\sigma(\Delta t)<2.5~{\rm ps}$, where $\sigma(\Delta t)$ is the error
on $\dt$ estimated on a per-event basis.

Discrimination between $\rhopi$ and $\rhok$ signal events is 
accomplished using the Cherenkov angle measurement
from the DIRC. Therefore, only $\rhoX$~candidates with bachelor track 
inside the geometrical acceptance of the DIRC 
are considered. The number of photons in the DIRC associated 
with the bachelor track must be greater than 5. 
In addition, the Cherenkov angle  $\theta_{Ch}$ of the bachelor track 
is required to be inconsistent with the proton hypothesis. 
Finally, we reject events where the bachelor track is inconsistent by 
more than $4\sigma$ with both the pion and kaon hypotheses.

\begin{figure}[t]
  \epsfxsize10cm
  \centerline{\epsffile{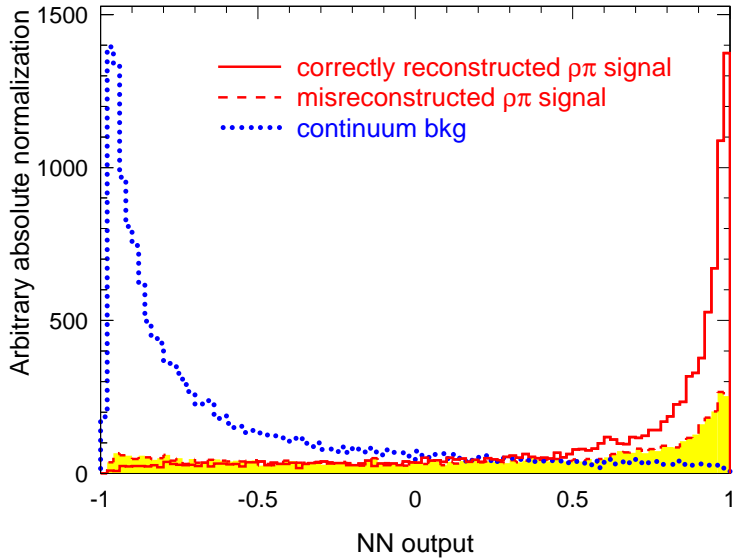}}
  \caption[.]{\label{fig:NN} Distributions of the NN output for correctly 
    reconstructed $\rhopi$ signal, misreconstructed $\rhopi$ signal and 
    continuum background.}
\end{figure}

Continuum $q\bar{q}$ (where $q = u,d,s,c$) events represent the
dominant background source for charmless $B$ decays.  To enhance
discrimination between signal and continuum background, we use a
neural network (NN) that combines four discriminating variables: two
kinematic variables related to the reconstructed $\rho^\pm$ candidate,
the $\rho$ mass and $\cshelrho$, and two event shape variables, $L_0 =
\sum_{i} p^*_i$ and $L_2 = \sum_{i} p^*_i \times|{\rm
cos}(\theta^*_{T_B,i})|^2$, where $p^*_i$ is the momentum of track~$i$
belonging to the rest of the event in the CM frame
 and $\theta^*_{T_B,i}$ is the angle
between the momentum of track $i$ and the $B$ thrust axis $T_B$
in the CM frame.
Optimization and training of the \NN\ is performed using off-resonance
data contained in the signal region, to reduce residual correlations
of the \NN\ with the kinematic variables used in the maximum
likelihood fit.  In addition, the Monte Carlo signal training sample,
generated with a GEANT4-based Monte Carlo simulation~\cite{GEANT4},
only consists of correctly reconstructed signal events to increase the
discrimination against cross-feed from other decay modes of the~$B$.
The distributions of the NN output for correctly reconstructed
$\rhopi$~events, $\rhopi$~events with a misreconstructed $\pi^0$, and 
continuum background are shown in
Fig.~\ref{fig:NN}.  A cut is applied on the NN output at 0.54 in order
to reduce the number of continuum events entering the likelihood
analysis. In addition, the NN output is used as a discriminating
variable in the likelihood.  Approximately 37\% of the events have
more than one candidate passing this selection.  In this case, we
choose the candidate with the reconstructed $\pi^0$ invariant mass
closest to the nominal $\pi^0$ mass~\cite{PDG2002}.

After all selection criteria have been applied, a total of 21172 events enter the
likelihood fit.
For $86\%$ of the neutral $B$ decays to $\rho^\pm\pi^\mp$ that pass the event selection, 
the charge of the $\rho$ is unambiguously
determined by the charge of the slowest track. If the lower-momentum track 
has a momentum below~$2.4$~\gevc, 
its charge is assigned to the~$\rho$.
This approach does not rely on the reconstruction of the $\pi^0$,
and hence provides a more robust way to assign the charge to the $\rho$ than criteria based on 
the reconstructed mass of $\rho$~candidates.
For the remaining events, the sign of the $\rho$ is that of the $\pi^{\pm}\pi^0$ combination
with invariant mass closest to the $\rho$ mass~\cite{PDG2002}. With
this procedure, only 5\% of the events are assigned an incorrect
charge.

To determine the flavor of the $\Bz_{\rm tag}$ meson we use the same
$B$-tagging algorithm used in the \babar\ $\stwob$
analysis~\cite{bib:BabarS2b}.  The algorithm relies on the correlation
between the flavor of the $b$~quark and the charge of the remaining
tracks in the event after removal of the tracks from the $B\to\rho h$
candidate.  We define five mutually exclusive tagging categories: {\tt
Lepton}, {\tt Kaon}, {\tt NT1}, {\tt NT2}, and {\tt Untagged}.  
{\tt Lepton} tags rely on primary electrons and muons from semileptonic 
$B$ decays, while {\tt Kaon} tags exploit the correlation in the process $b\to c\to s$ 
between the net kaon charge and the charge of the $b$~quark.  
The {\tt NT1}\,(more certain tags) and {\tt NT2}\,(less certain tags) categories 
are derived from a neural network that is sensitive 
to charge correlations between the parent \B\ and unidentified leptons and kaons, 
soft pions, or the charge and momentum of the track with the highest CM~momentum.
The addition of {\tt Untagged} events provides a larger sample for
measuring the charge asymmetries $\AcprhoK$ and $\Acprhopi$.

The quality of tagging is expressed in terms of the effective efficiency 
$Q = \sum_c \epsilon_c \left< D_c \right>^2$, where $\epsilon_c$ is the fraction of events tagged in 
category $c$ and the dilution $\left< D_c \right> = 1-2 \left< w_c \right>$ is 
related to the average mistag fraction $\left< w_c \right>$.
The mistag fraction $\left< w_c \right>$, the efficiency $\epsilon_c$, and   
the mistag difference $\Delta\mistag_c=\mistag_c-\overline\mistag_c$, 
where $\mistag_c$, $\overline\mistag_c$ are the mistag probabilities for $B^0$ and $\bar{B^0}$, 
are measured for each tagging category~$c$ with a large data sample of 
fully reconstructed neutral $B$ decays to $D^{(*)-}x^+\,(x^+ = \pip, \rho^+, a_1^+)$ 
and $\jpsi K^{*0}\,(K^{*0}\to\Kp\pim)$ flavor eigenstates~\cite{bib:BabarS2b}. 
The tagging quality factor $Q$  is found to be $(26.2\pm 0.7)\%$ (see Table~\ref{tab:tagging}).
We use these measurements of the tagging efficiencies and dilutions for $\rhopi$ and $\rho K$~signal,
and we float separate continuum background event yields for each category in the maximum 
likelihood fit.

\begin{table}[!tbp]
\caption{Tagging efficiency $\epsilon_c$, average dilution 
$\left< D_c \right> = 1 - 2\left< w_c \right>$, dilution difference $\diffD_c = -2 \Delta w_c$, 
and effective tagging efficiency $Q_c$ for signal events in each tagging category.  The
values are measured with fully reconstructed neutral $B$ decays.}
\smallskip
\begin{center}
\begin{tabular}{ccccc} \hline\hline
Category & $\epsilon_c\,(\%)$ & $\left< D_c \right>\,(\%)$ & $\diffD_c\,(\%)$ & $Q_c\,(\%)$ \rule[-2mm]{0mm}{6mm} \\\hline
{\tt Lepton}   & $10.7\pm 0.2$ & $83.8 \pm 1.6$ & $-0.4  \pm 2.4$ & $7.5\pm  0.3$ \rule[-1.5mm]{0mm}{5mm}\\
{\tt Kaon}     & $34.8\pm 0.3$ & $66.0 \pm 1.2$ & $ 3.6  \pm 1.6$ & $15.1\pm 0.5$ \rule[-1.5mm]{0mm}{4mm}\\
{\tt NT1}      & $7.7 \pm 0.2$ & $58.4 \pm 2.4$ & $-1.4  \pm 3.6$ & $2.6\pm  0.2$ \rule[-1.5mm]{0mm}{1.5mm}\\
{\tt NT2}      & $14.1\pm 0.2$ & $25.4 \pm 2.0$ & $ 7.2  \pm 3.0$ & $0.9\pm  0.2$ \rule[-1.5mm]{0mm}{1.5mm}\\
{\tt Untagged} & $32.7\pm 0.4$ & -- 	  & --            & --          \rule[-1.5mm]{0mm}{1.5mm}\\ \hline
Total $Q$ & & & & $26.2\pm 0.7$ \rule[-2mm]{0mm}{6mm} \\\hline\hline
\end{tabular}
\end{center}
\label{tab:tagging}
\end{table}

%% file: BBackground.tex
\section{\boldmath$B$-related Backgrounds}
\label{sec:BBackground}
We use a Monte Carlo simulation to study the potential cross-feed from other charm
and charmless $B$~decays starting from a list of more than
$80$~charmless decay modes to two-body, three-body and
four-body final states, and an inclusive Monte Carlo simulation of $B \to {\rm charm}$ decays. 
We estimate the number of events passing the event selection criteria
using the selection efficiency from Monte Carlo and either measured branching 
ratios~\cite{bib:Bbkg} or upper limits where available, or 
estimates based on related measured decay modes. We identify the $20$~charmless modes 
that have more than one event entering the final sample.

These modes are grouped into seven classes for which the discriminating variables
have similar distributions. For each of the seven classes, a correction term is 
introduced in the likelihood, with a fixed number of events. Two additional classes for 
$\Bu \to {\rm charm}$ and $\Bz \to {\rm charm}$ decays are also included in the 
$B$-background model, which is summarized in Table~\ref{tab:BbkgClasses}.

\begin{figure}[htb]
  \epsfxsize16cm
  \centerline{\epsffile{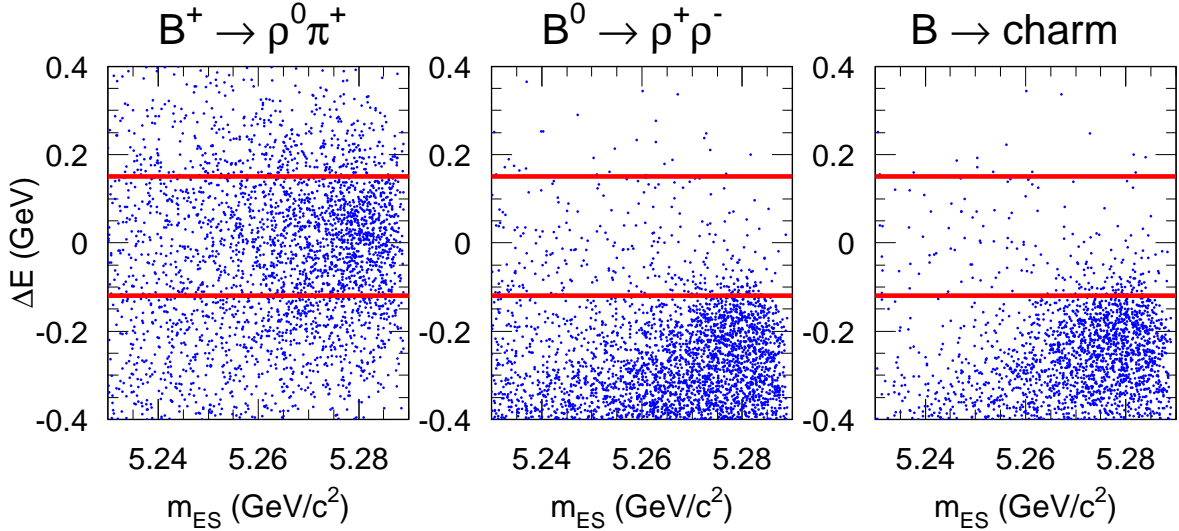}}
  \caption[.]{\label{fig:BbkgPDFs} $\de$-$\mes$ plane for three $B$-related backgrounds 
(Monte Carlo simulation): 
$B^+ \rar \rho^0\pi^+$, $B^0 \rar \rho^+\rho^-$ and $B \rightarrow$ charm. The lines
indicate the cut on~$\de$ applied in the analysis;
we require $-0.12<\de<0.15~\gev$. It removes most of the four-body
$B$-background modes (such as $B^0 \rar \rho^+\rho^-$) and $B \to {\rm charm}$ modes.}
\end{figure}

Like the selection efficiencies, the shapes of the distributions of the discriminating 
variables are obtained from Monte Carlo
simulations. Figure~\ref{fig:BbkgPDFs} shows the $\de$-$\mes$ planes
for three main $B$-related backgrounds: 
$\B^+\to\rho^0\pi^+$, $\Bz\to\rho^+\rho^-$, and $\B \to {\rm charm}$.
The charmless $B$-background NN output and $m_{ES}$
distributions are signal-like, and the $\Delta E$ variable
discriminates between two-body ($\Delta E>0$), three-body ($\Delta E$ 
peaking around 0) and four-body ($\Delta E<0$) modes.  The $m_{ES}$ and 
$\Delta E$~distributions for $B \to {\rm charm}$ background have shapes similar to 
the continuum distributions.

For charged $B$-backgrounds, the $\Delta t$~distribution is modeled as
\begin{eqnarray}
\nonumber
g_{B^0_{\rm tag, c}}^{\rho^{\pm}h}(\Delta t) &=& 
\frac{1}{4\tau}\left[ 1\pm A_{h} (1-2w^{\pm}_{c}) \right] e^{-|\Deltat|/\tau}, \\
g_{\bar{B}^0_{\rm tag, c}}^{\rho^{\pm}h}(\Delta t) &=& 
\frac{1}{4\tau}\left[ 1\mp A_{h} (1-2w^{\pm}_{c}) \right] e^{-|\Deltat|/\tau},
\end{eqnarray}
where  $A_{h}$ is the asymmetry between the number of $\rho^+ h^-$ and $\rho^- h^+$ 
candidates for a given flavor tag and is extracted 
from Monte Carlo, and $w^{\pm}_{c}$ is the mistag fraction for tagging 
category $c$, measured in data using a sample of fully reconstructed 
charged $B$ decays to~$D^{*0}\pi^+$ for which we assume direct \CP\ conservation.

For neutral $B$-backgrounds, the $\Delta t$ distribution is parametrized as for signal, where
$A_{CP}^h = S_h = \Delta S_h = C_h = 0$, and $\Delta C_h$ is 
computed from Monte Carlo to take into account possible correlations between the reconstructed 
$\rho$~charge and the flavor tag. We do not model \CP~violation for the $B$-background
in the nominal fit. The corresponding systematic uncertainties in our measurements are 
discussed in Sec.~\ref{sec:Systematics}.

\begin{table}[htb]
\begin{center}
\caption{$B$-background modes retained in the maximum likelihood fit, classified into nine
categories. The number of expected events contributing to the $\rhopi$ and $\rhok$ PDFs,
scaled to $80.8\,\invfb$, and integrated over the full fit region,
is reported in the 2nd and 3rd columns. The 4th and 5th columns give the assumed parameter values for
the $\Delta t$ distributions (see text). The last column gives the
branching ratio (in units of $10^{-6}$)
if measured, or the estimated range if not (these cases are indicated by the symbol~$^*$).}
\vspace{0.3cm}
\label{tab:BbkgClasses}
\begin{tabular}{lccccc}
\hline
\hline
 Charged Mode & $N_{\rm exp}^\pi$ & $N_{\rm exp}^K$ & $A_\pi$ & $A_K$ & (Br $\pm$ error) 
$(10^{-6})$\\\hline
$B^+ \rar K^{*+}(K^+\pi^0)\rho^0$ & 0.41 & 2.16 & 1 & -1 & $3\,-\,13^*$\\
$B^+ \rar K^{*0}(K^+\pi^-)\rho^+$ & 0.06 & 7.23 & 1 & -1 & $10\,-\,40^*$\\
$B^+ \rar \rho^+\rho^0$ & 15.17 & 0 & 0.22 & -- & $10\,-\,20^*$ \\
$B^+ \rar \eta'(\rho^0\gamma)K^+$ & 0.19 & 7.78 & 1 & -1 & 22.1 $\pm$ 2.1\\
$B^+ \rar \eta'(\rho^0\gamma)\pi^+$ & 1.53 & 0 & -1 & -- & $1\,-\,5^*$\\ \hline
$B^+ \rar \rho^+\pi^0$ & 23.22 & 0 & -1 & -- &  $7\,-\,23^*$ \\
$B^+ \rar \rho^0 K^+$ & 2.05 & 21.20 & 0.80 & -1 &  8.4 $\pm$ 4.0 \\
$B^+ \rar \rho^0\pi^+$ & 36.51 & 0 & -0.46 & -- & 9.7 $\pm$ 3.2 \\
$B^+ \rar K^0_S \pi^+$ & 8.24 & 0 & -0.76 & -- & 8.7 $\pm$ 1.3 \\
$B^+ \rar K^+f_0(\pi^+\pi^-)$ & 1.69 & 15.46 & 1 & -1 & 11.7 $\pm$ 4.0 \\ \hline
$B^+ \rar K^+\pi^0$ & 0.17 & 13.50 & 0 & -1 & 11.6 $\pm$ 1.5 \\
$B^+ \rar \pi^+\pi^0$ & 5.09 & 0 & -1 & 0 & 5.9 $\pm$ 1.4\\ \hline
$B^+ \rar \mathrm{charm}$ & 195.0 & 31.8 & 0 & -0.46 & \\\hline
\hline
 Neutral Mode & $N_{\rm exp}^\pi$ & $N_{\rm exp}^K$ & $\Delta C_\pi$ & $\Delta C_K$ & (Br $\pm$ error) $(10^{-6})$ \\\hline
$B^0 \rar K^{*+}(K^0_S\pi^+)\pi^-$ & 3.38 & 0 & 1 & -- & 8.7 $\pm$ 3.0\\
$B^0 \rar K^{*+}(K^+\pi^0)\rho^-$ & 0.72 & 5.57 & 1 & -1 & $0\,-\,20^*$ \\\hline
$B^0 \rar \rho^+\rho^-$ & 87.64 & 0 & 0 &  -- & $40\,-\,100^*$\\
$B^0 \rar \rho^0\rho^0$ & 1.03 & 0 & 0 & -- & $0\,-\,3^*$\\\hline
$B^0 \rar a_1^+(\rho^0\pi^+)\pi^-$ & 9.43 & 0 & 0 &  -- & $28\,-\,48^*$\\
$B^0 \rar K^{*0}(K^+\pi^-)\pi^0$ & 0 & 6.62 & -- & -1 & $0\,-\,6^*$ \\\hline
$B^0 \rar K^{*+}(K^+\pi^0)\pi^-$ & 20.92 & 12.97 & 0.85 & -1 & 8.7 $\pm$ 3.0 \\
$B^0 \rar K^+\pi^-$ & 1.87 & 2.17 & 1 & -1 & 18.5 $\pm$ 1.0\\\hline
$B^0 \rar \mathrm{charm}$ & 121.7 & 13.7 & 0 & 0 & \\
\hline
\hline
\end{tabular}
\end{center}
\end{table}

%% file: ML.tex
\section{The Maximum Likelihood Fit}
\label{sec:ML}

\subsection{The Likelihood}
The yields and the other \CP\ and non-\CP\ observables are determined by
minimizing the quantity $-2\ln{\cal L}$, where ${\cal L}$ is the total
extended likelihood defined over all tagging categories $c$ as
\begin{equation}
	{\cal L} = \prod_{c=1}^{5} e^{-N^{\prime}_c}\prod_{i=1}^{N_c} {\cal L}_{i,\,c}~,
\end{equation}
with $N^{\prime}_c$, the number of events expected in category $c$ and
${\cal L}_{i,\,c}$ is the likelihood computed for event~$i$.

The sample is assumed to consist of signal, continuum background and $B$-background 
components where the bachelor track can be a pion or a kaon. 
The variables $\mes$, $\de$ and \NN\ output 
discriminate signal from background, while
the Cherenkov angle $\theta_{Ch}$ and, to a lesser extent, $\de$
constrain the relative amount of $\rhopi$ and $\rhok$. The variable~$\deltat$
allows the measurement of the parameters in the description of the 
$\Bz(\Bzb) \to \rhopi$ decay-time dependence and provides additional 
background rejection. The
likelihood ${\cal L}_{i,\,c}$ for event~$i$ in tagging
category $c$ is the sum of the probability density functions~(PDF) over
all components, weighted by the expected yields for each component,
\begin{eqnarray}
\label{eq:pdfsum}
{\cal L}_{i,\,c} &=& N_{\rho\pi} \epsilon_c {\cal P}_{i,\,c}^{\rho\pi}
+ N_{\rho K} \epsilon_c{\cal P}_{i,\,c}^{\rho K}
+\: N^c_{q\rho\pi} {\cal P}_{i,\,c}^{q\rho\pi} 
+ N^c_{q\rho K} {\cal P}_{i,\,c}^{q\rho K} 
 +\: {\cal L}^{\B, \pi}_{i,\,c} + {\cal L}^{\B, K}_{i,\,c} \; ,
\end{eqnarray}
where
\begin{itemize}
 \item 	$N_{\rho h}$ is the number of signal events of type ${\rho h}$ in 
	the entire sample ($h = \pi , K$).
 \item 	$\epsilon_c$ is the fraction of signal events that are 
       	tagged in category $c$ (given in Table~\ref{tab:tagging}).
 \item 	$N_{q\rho h}^c$ is the number of continuum background events with 
        bachelor track of type~$h$ that are tagged in category~$c$.
 \item ${\cal P}_c^{\rho h} = {\cal P}^{\rho h}(\mes)\cdot {\cal
	P}^{\rho h} (\de)\cdot {\cal P}^{\rho h}(\NN)\cdot {\cal
	P}^{\rho h} (\theta_{Ch})\cdot {\cal P}_c^{\rho h}(\deltat)$ is
	the PDF for signal events. ${\cal P}_c^{\rho h}(\deltat)$
	contains the measured physics quantities defined in
	Eq.~\ref{eq:thTime} diluted by the effects of mistagging 
	and the~$\deltat$ resolution.
 
 \item   ${\cal P}_{i,\,c}^{q\rhoX}$ is the
	PDF for continuum background events with bachelor track of type~$h$.

 \item  ${\cal L}^{\B,\pi}_c$ and ${\cal L}^{\B, K}_c$ are the 
        \B-background contributions where the bachelor track is a true pion 
        and a true kaon, respectively (see Sec.~\ref{sec:BBackground}).
\end{itemize}
Due to the relatively large number of low-energy photon candidates in 
$\pi^0$ reconstruction, the decay $\rho^{\pm}\pi^{\mp}
\rightarrow \pi^+\pi^-\pi^0$ can be misreconstructed. These
misreconstructed events have different shapes than correctly 
reconstructed signal for the distributions of the variables $\mes$, $\de$ 
and \NN\ output. Additionally, in some cases the assignment of the $\rho$
charge may be wrong. These effects are taken into account by splitting
the signal PDFs into three parts: true signal events that are correctly reconstructed, 
misreconstructed true signal events with right-sign charge, and misreconstructed true 
signal events with wrong-sign charge. The fractions of the three species of signal 
events are extracted from the Monte Carlo.

\subsection{The Probability Density Functions}

\begin{itemize}
\item {\boldmath$\mes$}\\
	The distribution for correctly reconstructed signal is parametrized using a 
        Gaussian with a power law tail on the low side, 
	where the mean is free to vary in the likelihood fit.
	The continuum background is parametrized using 
	an ARGUS function~\cite{bib:argusshape} with a floating shape parameter.
	
\item	{\boldmath$\de$}\\
	The distribution for correctly reconstructed signal is parametrized using the sum of 
        two Gaussians, while the distributions for misreconstructed signal (both for 
        right-sign and wrong-sign $\rho$ charges)
	are modeled with simple Gaussians. The mean of the core Gaussian 
	for the correctly reconstructed signal is floated in the maximum likelihood fit, 
	in order to be less sensitive 
	to the energy calibration  for $\pi^0$.	Continuum background is modeled by a linear 
	function.
	
\item	{\bf \boldmath$\NN$ output}\\
	The NN output PDFs for correctly reconstructed and for misreconstructed signal 
        events are determined with the
	Monte Carlo. A small discrepancy is observed between the NN output distributions
	from Monte Carlo and from a data control sample of fully reconstructed 
        $B^{0} \rightarrow D^{-} \rho^{+}$ decays, and is propagated to the systematic 
        error. The continuum PDF for the NN output is determined with the
	off-resonance data.
\end{itemize}

For the above three variables, the PDFs for correctly reconstructed $\btorhc$ 
decays and for misreconstructed $\btorhc$ decays are obtained 
from Monte Carlo after applying all selection cuts. The parameters for
the continuum PDFs are either determined from off-resonance data,
or left free to vary in the final fit to the on-resonance data sample.

\begin{itemize}
\item 	{\bf \boldmath$\dt$}\\
	The resolution function for correctly reconstructed and misreconstructed 
        signal events is a sum of three Gaussians, identical 
	to the one described in Ref.~\cite{bib:BabarS2b}, with parameters determined from a 
	fit to a large data sample of fully reconstructed neutral $B$ decays to 
        $D^{(*)-}x^+\,(x^+ = \pip, \rho^+, a_1^+)$ and $\jpsi K^{*0}\,(K^{*0}\to\Kp\pim)$. 
	The continuum $\deltat$ distribution is parametrized using a triple Gaussian with 
	a common mean and three distinct widths that scale the $\dt$ per-event error. The six 
	parameters describing the shape of the the $\dt$ continuum PDF are free to vary in the 
        maximum likelihood fit. For each tagging category and bachelor hypothesis ($\pi$~or~$K$),
        a parameter similar to~$A_{h}$ for $B$-related backgrounds (see Sec.~\ref{sec:BBackground}) 
        is introduced to take into account the correlations between the charge of the 
        $\rho$~candidate and the flavor tag. The values of these parameters are determined 
        using on-peak data.

\item 	{\bf \boldmath Particle identification}\\
	The identification of the bachelor track as a pion or a
	kaon is accomplished with the Cherenkov angle measurement from the
	DIRC. We construct two Gaussian PDFs from the difference between
	measured and expected values of $\theta_{Ch}$ for the pion or kaon
	hypothesis, normalized by the resolution.  The DIRC performance is
	parametrized using a data sample of $D^{*+}\to\Dz\pip$, $\Dz\to
	\Km\pip$ decays.  Within the statistical precision of the control
	sample, we find a similar response for positive and negative tracks and
	use a single parametrization for both.

\end{itemize}

%% file: Results.tex
\section{Results}
\label{sec:Results}

We find $413^{+34}_{-33}$~(stat) $\rhopi$ and $147^{+22}_{-21}$~(stat) $\rho K$ events 
in our data sample and we measure the \CP~parameters
\begin{eqnarray}
\label{eq:res1}
 \AcprhoK = 0.19 \pm 0.14 \,\,\,{\rm(stat)},  & &
 \Acprhopi = -0.22 \pm 0.08\,\,\,{\rm(stat)},\nonumber \\ 
 C_{\rho\pi} = 0.45^{+0.18}_{-0.19}\,\,\,{\rm(stat)},  & &
 S_{\rho\pi} = 0.16 \pm 0.25\,\,\,{\rm(stat)}.   
\end{eqnarray}
The two other observables in the decay rates (Eq.~\ref{eq:thTime}) are measured to be
\begin{eqnarray}
\label{eq:res2}
 \dC_{\rho\pi} \, = \, 0.38^{+0.19}_{-0.20}\,\,\,{\rm(stat)}, &&
\dS_{\rho\pi} =0.15 \pm 0.26 \,\,\,{\rm(stat)}.
\end{eqnarray}
The correlations between these parameters are summarized in 
Table~\ref{tab:correlation}.\\
Alternatively, the results on direct \CP~violation can be expressed using the asymmetries
\begin{eqnarray}
\label{eq:Asy1}
A_{+-} & = & \frac{N({\Bzb_{\rhopi}} \to {\rho^+\pi^-})
         - N({\Bz_{\rhopi}} \to {\rho^-\pi^+})}
        {N({\Bzb_{\rhopi}} \to {\rho^+\pi^-})
         + N({\Bz_{\rhopi}} \to {\rho^-\pi^+})} 
         =  \frac{\Acprhopi  - C_{\rho\pi} - \Acprhopi \cdot \dC_{\rho\pi}}
        {1 - \dC_{\rho\pi} -  \Acprhopi \cdot C_{\rho\pi} } \; ,\\
\label{eq:Asy2}
A_{-+} & = & \frac{N({\Bzb_{\rhopi}} \to {\rho^-\pi^+})
         - N({\Bz_{\rhopi}} \to {\rho^+\pi^-})}
        {N({\Bzb_{\rhopi}} \to {\rho^-\pi^+})
         + N({\Bz_{\rhopi}} \to {\rho^+\pi^-})}
        =  -\frac{\Acprhopi  + C_{\rho\pi} + \Acprhopi \cdot \dC_{\rho\pi}}
        {1 + \dC_{\rho\pi} +  \Acprhopi \cdot C_{\rho\pi} }\; .
\end{eqnarray}
In the decays $\Bzb_{\rhopi} \to \rho^+\pi^-$ and $\Bz_{\rhopi} \to \rho^-\pi^+$ the spectator quark is
involved in the formation of the $\rho$~meson. These two decay modes are
related to the direct \CP~asymmetry~$A_{+-}$ according to Eq.~\ref{eq:Asy1}.
Similarly in Eq.~\ref{eq:Asy2}, we probe direct \CP~violation through the asymmetry~$A_{-+}$
using the decays $\Bzb_{\rhopi} \to \rho^-\pi^+$
and $\Bz_{\rhopi} \to \rho^+\pi^-$.
In this case the $\pi$ meson is formed from the spectator quark.
From the above fitted values (Eq.~\ref{eq:res1}
and Eq.~\ref{eq:res2}) and their correlation matrix 
(Table~\ref{tab:correlation}) we obtain
\begin{eqnarray*}
A_{+-} \, = \, -0.82\pm 0.31\,\,\,{\rm(stat)}, &&
A_{-+} \, = \, -0.11\pm 0.16 \,\,\,{\rm(stat)}.
\end{eqnarray*}
 
\begin{figure}[h]
  \epsfysize9.cm
  \centerline{\epsffile{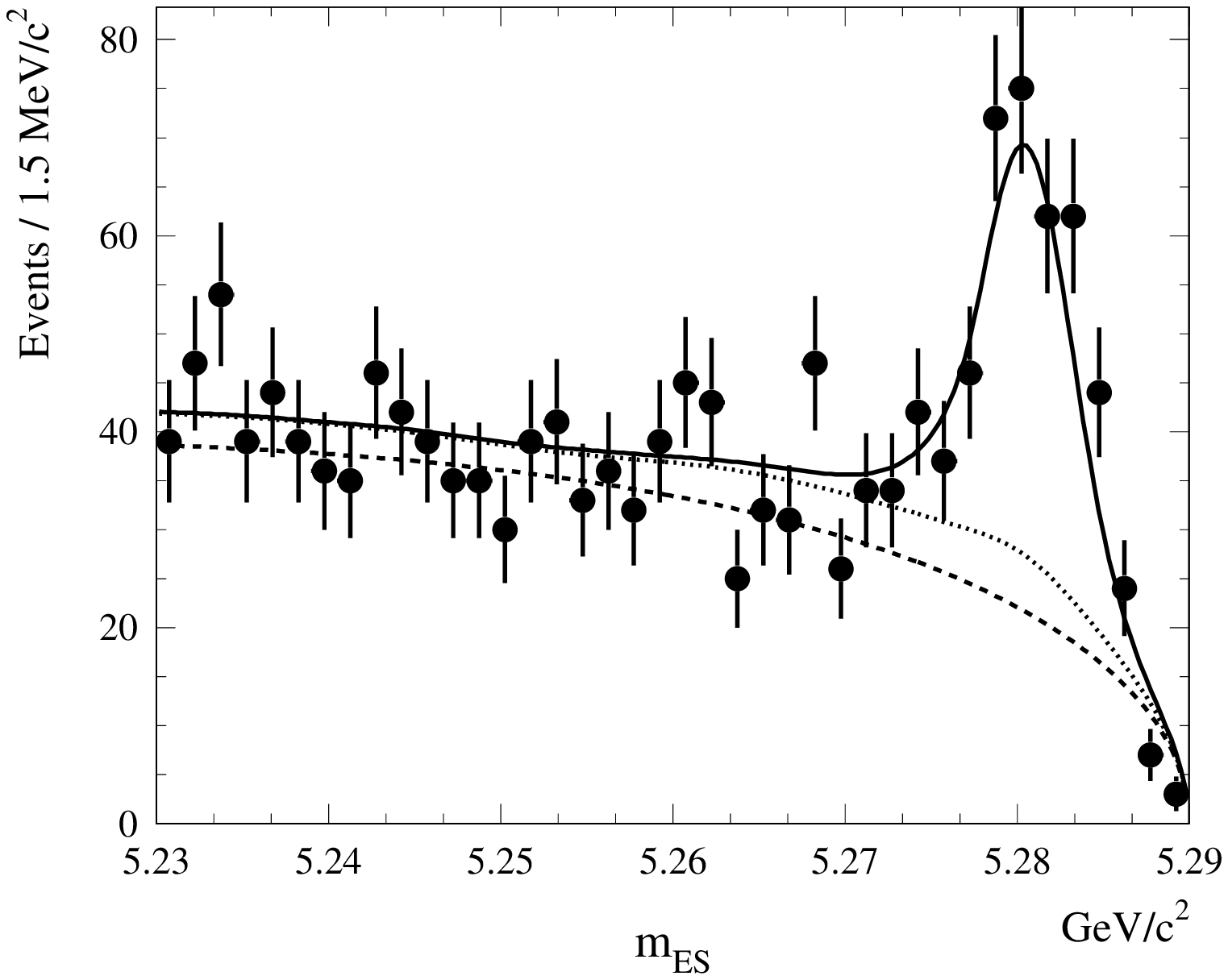}}
  \centerline{\epsffile{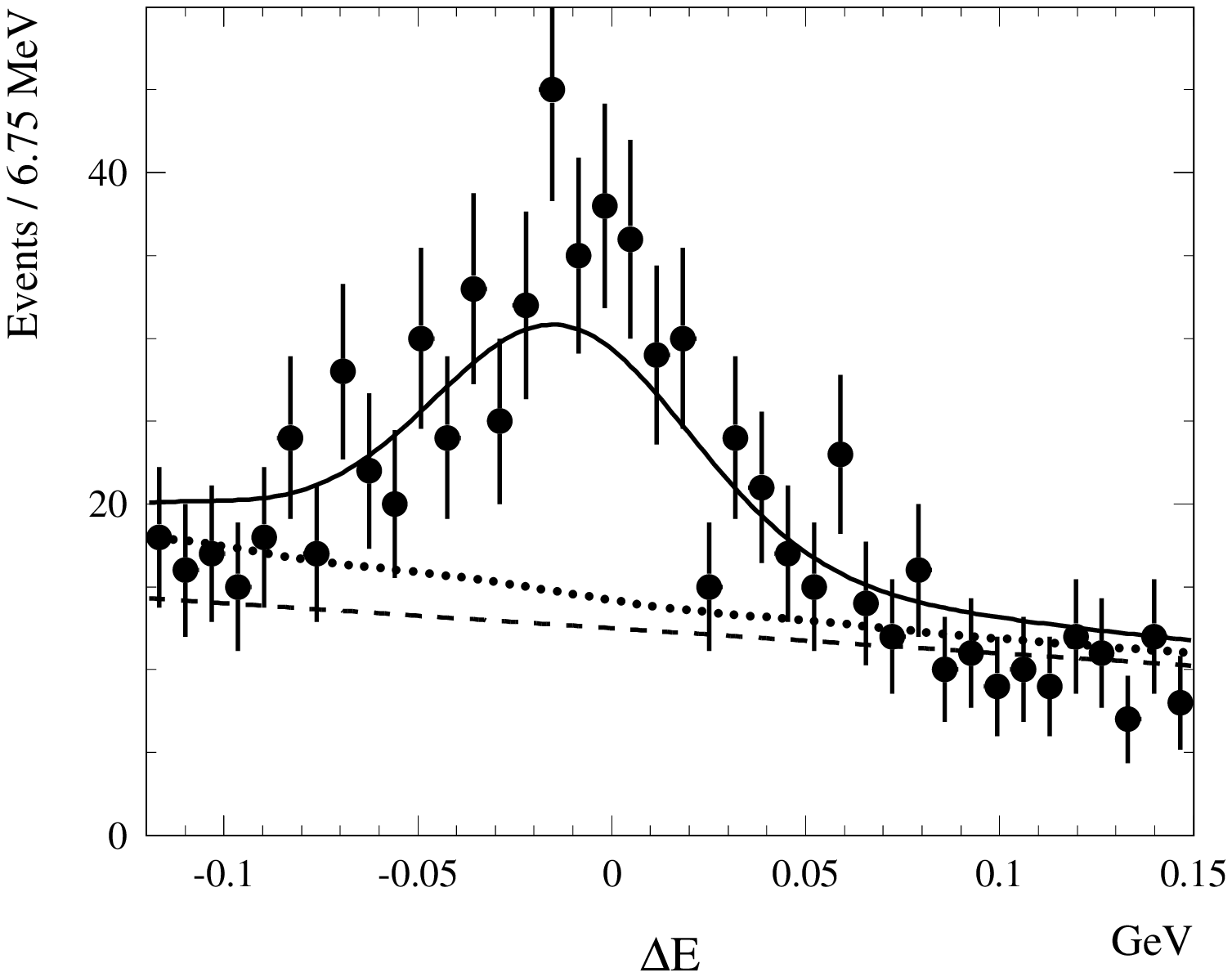}}
  \caption[.]{Distributions of~\mes and~$\Delta E$ for samples enhanced 
in $\rho\pi$~signal using cuts on likelihood ratios. The solid curve 
represents a projection of the maximum likelihood 
fit result. The dashed curve 
represents the contribution from continuum events ($\rho\pi$~and $\rho K$~candidates combined), 
and the dotted line indicates the combined contributions from continuum events and $B$-related backgrounds, including~$\rho K$.}
\label{fig:ProjMesDE}
\end{figure}

\begin{figure}[htb]
  \begin{center}
    \includegraphics[width=0.8\textwidth]{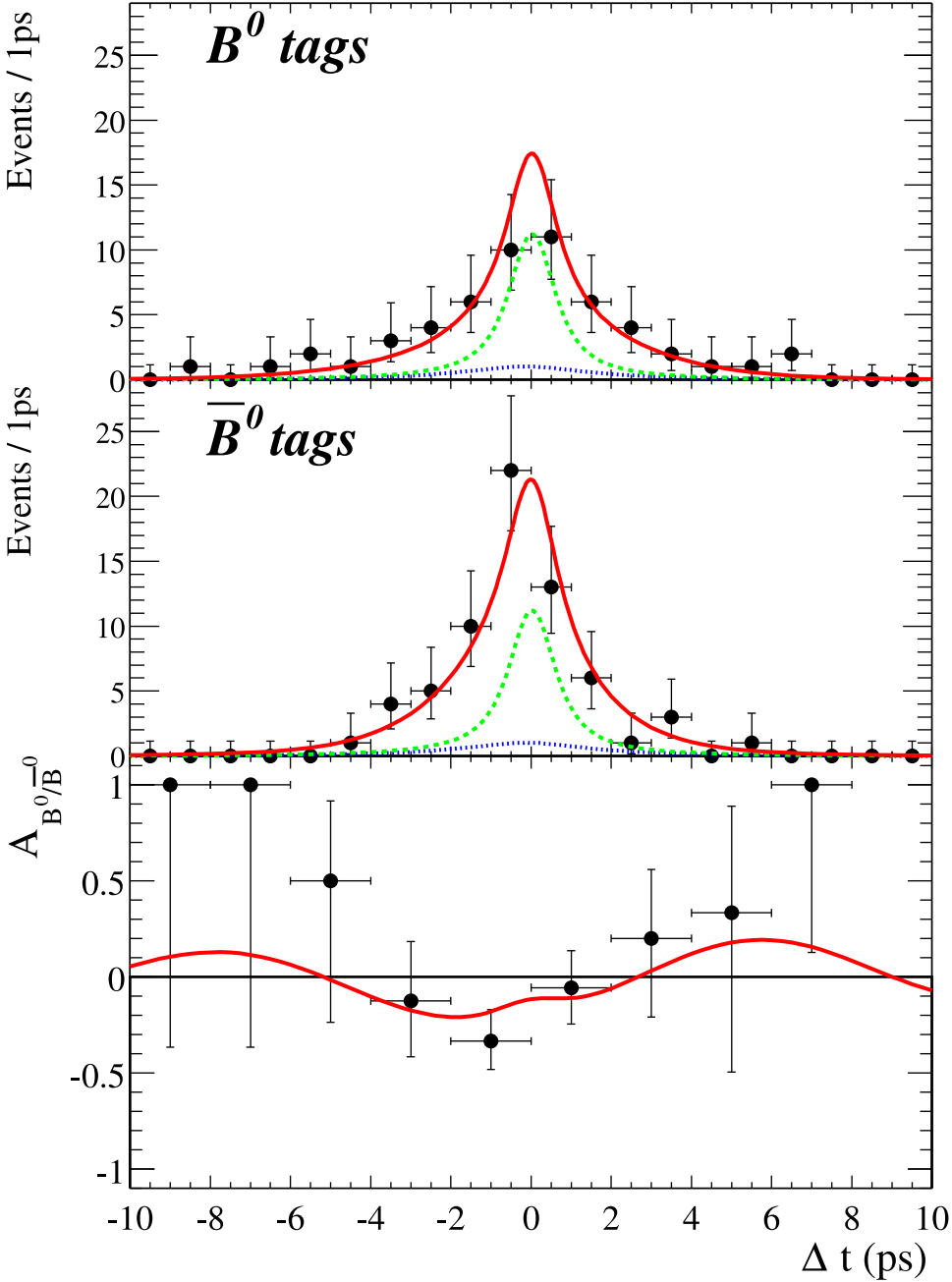}
  \end{center}

\caption{Time distribution and asymmetry ($A_{\Bz/\Bzb}$ in
  Eq.~\ref{eq:Assym}) for
  $\Bz_{\rm tag}$ and $\Bzb_{\rm tag}$ decaying to $\rho\pi$, in the
  {\tt Lepton} and {\tt Kaon} categories. The sample was enriched in
  signal events by applying a cut on the signal-to-continuum
  likelihood ratio. The solid curve
  is a likelihood projection of the result of the full fit, and is
  normalized to the expected number of events according to that
  fit (71 signal events, 36 continuum background events and 10 \B
  background events). The dotted line is the contribution from $B$-related backgrounds and the
  dashed line is the total \B and continuum background contribution.
  The depression around zero in the asymmetry plot is due to the relatively 
  large fraction of continuum events in this region of~\deltat. The 
  non-zero central value for the \CP\ parameter $\S_{\rhopi}$ induces a small contribution 
  to the asymmetry that is odd in~\deltat.}
\label{fig:asymCS} 
\end{figure}

\begin{figure}[htb]
  \begin{center}
    \includegraphics[width=0.8\textwidth]{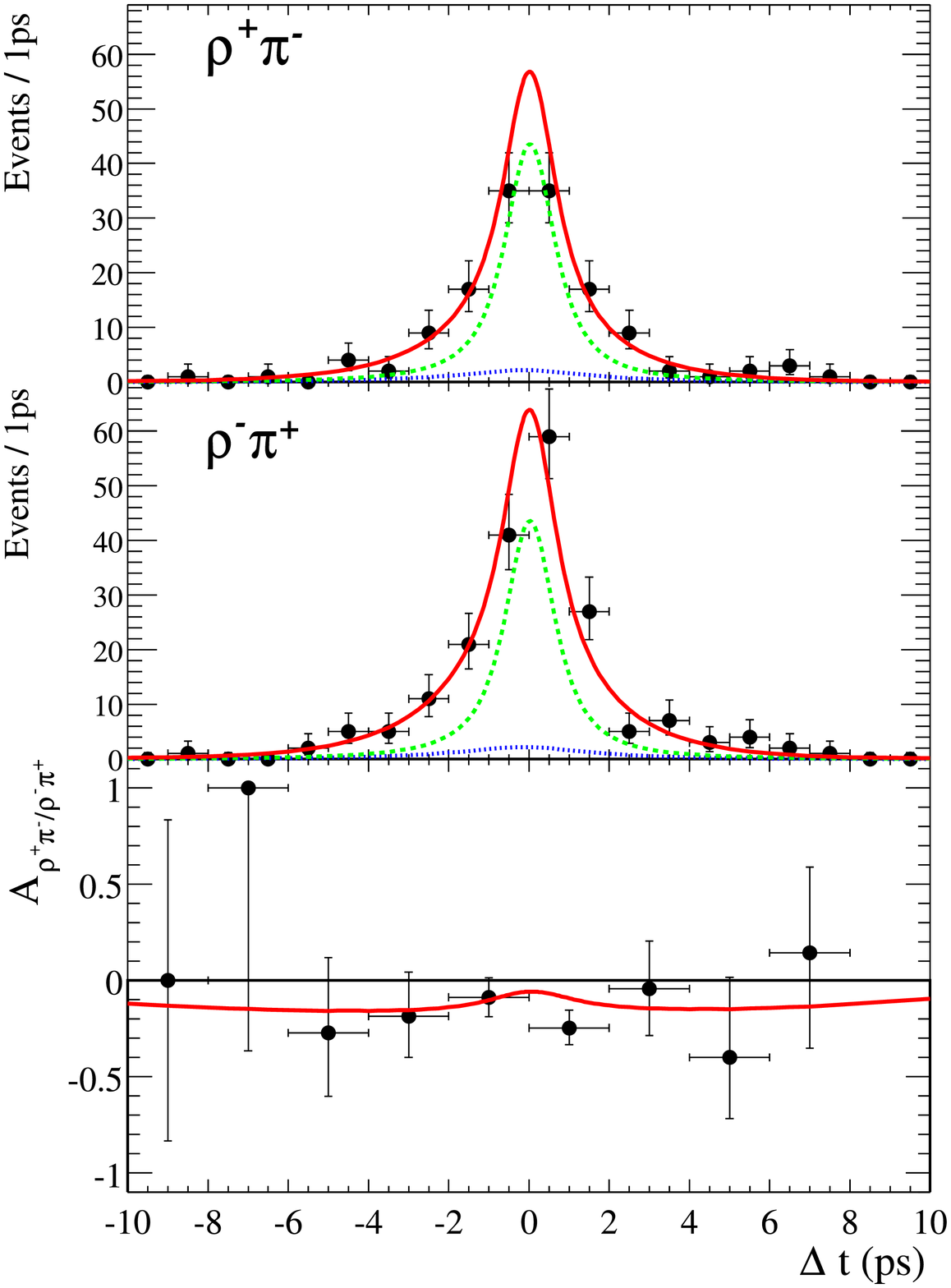}
  \end{center}

\caption{Time distribution and   asymmetry  $A_{\rho^+\pi^-/\rho^-\pi^+}$ 
 between $\rho^+\pi^-$ and $\rho^-\pi^+$ for all the tagging categories. 
 The sample was enriched in
  signal events by applying a cut on the signal-to-continuum
  likelihood ratio. The solid curve
  is a likelihood projection of the result of the full fit, and is
  normalized to the expected number of events according to that
  fit (156 signal events, 157 continuum background events and 21 \B
  background events). The dotted line is the contribution from $B$-related backgrounds and the
  dashed line is the total \B and continuum background contribution.
  The depression around zero in the asymmetry plot is due to continuum
  dilution. In the absence of continuum background, 
  the asymmetry curve would be flat and equal
  to $\Acprhopi$ defined in Eq.~\ref{eq:thTime}.}
\label{fig:asymACP} 
\end{figure}

Figure~\ref{fig:ProjMesDE} shows the distributions of $\mes$ and $\de$ for data samples that 
are enhanced in signal using cuts on the signal-to-continuum likelihood ratio of the other
discriminating variables.
Figures~\ref{fig:asymCS} and~\ref{fig:asymACP} show respectively 
the time-dependent asymmetry $A_{\Bz/\Bzb}$ (see Eq.~\ref{eq:Assym})
between $\Bz_{\rm tag}$ and $\Bzb_{\rm tag}$ events 
in the {\tt Lepton} and {\tt Kaon} categories,
and the time-dependent asymmetry $A_{\rho^+\pi^-/\rho^-\pi^+}$ 
 between $\rho^+\pi^-$ and $\rho^-\pi^+$ for all the tagging categories,
 after a cut on the signal-to-continuum likelihood ratio of all discriminating variables except~\deltat.

\begin{table}[bt]
\caption{Correlation coefficients (in percent) between the six parameters $\AcprhoK$, 
$\Acprhopi$, $C_{\rho\pi}$, 
$\dC_{\rho\pi}$, $S_{\rho\pi}$ and $\dS_{\rho\pi}$ obtained for a data sample
of 88 million $B\Bbar$ pairs. 
The global correlation coefficient is the largest 
correlation between the parameter in question and any linear combination 
of the other 28~free parameters in the likelihood fit.}
\begin{center}
\begin{tabular}{c|c|cccccc} \hline
\hline
& Global& $\Acprhopi$& $\AcprhoK$& $C_{\rho\pi}$& $\dC_{\rho\pi}$& $S_{\rho\pi}$& $\dS_{\rho\pi}$ \\
& correlation& & & & & &   \\ \hline
$A_{CP}^{\rho\pi}$& 15.6 & 100 & 3.4 & $-11.8$ & $-10.4$ & 0.6 & 0.5 \\
$A_{CP}^{\rho k}$& 6.3 & 3.4 & 100 & $-1.3$ & $-1.1$ & $-0.4$ & $-0.5$ \\
$C$& 28.8 & $-11.8$ & $-1.3$ & 100 & 23.9 & 9.2 & $-6.8$ \\
$\Delta C$& 28.3 & $-10.4$ & $-1.1$ & 23.9 & 100 & 6.9 & $-9.2$ \\
$S$& 24.9 & 0.6 & $-0.4$ & 9.2 & 6.9 & 100 & $-23.4$ \\
$\Delta S$& 25.1 & 0.5 & $-0.5$ & $-6.8$ & $-9.2$ & $-23.4$ & 100 \\ \hline
\hline
\end{tabular}
\end{center}

\label{tab:correlation}
\end{table}

As a validation of the $\deltat$ parametrization in data, we allow
$\tau$ and $\deltamd$ to vary in the fit. We find $\tau = (1.59\pm 0.12)\ps$ 
and $\deltamd = (0.51\pm 0.09)\ps^{-1}$, and the remaining free parameters 
are stable with respect to the nominal fit with fixed $\tau$ and $\deltamd$. 
When allowing $\dC_{\rho K}$ to vary, we find $\dC_{\rho K} = -1.15\pm0.34$, which
is consistent with the expected value of $-1$ used in the nominal fit.


\clearpage

%% file: Systematics.tex
\section{Systematic Uncertainties}
\label{sec:Systematics}

Several sources contribute to the systematic error on the 
measured observables. 
\begin{itemize}
\item {\bf \dm\ and $B$ lifetime}\\ 
This systematic error is estimated by varying the values of \dm\ and the $B$ 
lifetime with respect to their world average~\cite{PDG2002} by one standard deviation.
\item {\bf \dt\ resolution function}\\ 
Since the parameters of the time distribution for the continuum background are fitted in the data,
 the only contribution to the systematic uncertainty is due to the \dt\ resolution for signal.
We propagate the errors on each parameter (scale factors, biases, fraction of tails)
of the signal $\dt$ resolution function to the fit results. 
\item {\bf Tagging}\\ 
We propagate the uncertainties on the tagging dilutions and the tagging efficiencies
to the fit results.

\item {\bf \boldmath Fraction of misreconstructed signal events and misreconstruction of $\rho$~charge}\\  
Both the the fraction of misreconstructed signal events and the fraction of events with misreconstructed 
$\rho$ charge and are determined with the Monte Carlo. We assign a conservative 25\% uncertainty to these 
parameters and propagate this error to the fit results.
\item {\bf Fitting procedure}\\ 
We perform fits on large Monte Carlo samples of a weighted mixture
of $\rhopi/\rhok$ signal, as well as fits to a large number of toy Monte
Carlo samples of the same size as our data sample that contain realistic amounts of
both continuum and $B$-related backgrounds. The small biases observed in these two 
tests are added in quadrature and assigned as a systematic uncertainty.
\item {\bf \boldmath$\mes$, \boldmath$\de$ and NN output PDFs}\\
The continuum PDFs for $\mes$ and $\de$ are fitted to the data in the likelihood analysis.
The continuum PDF for the NN output is determined with the off-resonance data
and we propagate the statistical uncertainty due to the limited size of this sample
to the fit results. We evaluate the systematic uncertainties due to the signal PDFs with 
a large $B^{0} \rightarrow D^{-} \rho^{+}$ data control sample. The small  differences 
observed between the distribution shapes for Monte Carlo events and the distribution
shapes obtained from the data control 
sample are used as an estimate of the systematic uncertainty on the signal PDFs. These uncertainties 
are propagated to the fit results.

\item{\bf Particle identification}\\
We assume that the average position of the Cherenkov angle
$\langle\theta_{Ch}\rangle$ is known with a precision of 0.5 mrad and that
there is an uncertainty of 10\% on $\sigma(\theta_{Ch})$. We propagate these
conservative errors to our fit results.

\item{\bf \boldmath$B$-related backgrounds}\\
The number of events in the various modes entering our description
of the \B background are varied according to the error on their branching
ratio, if measured, or in the range indicated in Table~\ref{tab:BbkgClasses}.\\
The  parameters ($A_h$ and $\dC_h$) describing the correlation between the
tagging and the  $\rho$ charge assignment are varied
within a conservative range. For some of the major contaminations
($\rho^+\rho^-$, $\rho^0\pi^+$, and $B \to {\rm charm}$) conservative
ranges are extracted from various Monte Carlo studies. For other major
contaminations ($K^{*+}(K^+\pi^0)\pi^-$, $\rho^0 K$) a conservative
error equal to the correction itself is used. For the other modes, the
full range $\left[-1,+1 \right]$ is used for the systematic study.\\
All $B$-background modes can potentially exhibit direct \CP violation,
and a few of them can potentially exhibit \CP violation in the
interference between decay and mixing. Finally, for the neutral modes,
various physical phases may lead to a non-zero value of the
phenomenological parameter $\dS_h$. For the major contaminations listed
previously, a Monte Carlo study yields the maximum possible range for
the corresponding effective parameters. This range is then
used for the systematic study. For the other modes, the full range
$\left[-1,+1 \right]$ is used.
\end{itemize}

Table~\ref{tab:sys_table}  summarizes the various sources contributing 
to the systematic error on the measurements of the six parameters 
$\AcprhoK$, $\Acprhopi$, $C_{\rho\pi}$, $\dC_{\rho\pi}$, $S_{\rho\pi}$, 
and $\dS_{\rho\pi}$.  The main source 
of systematic error arises from the uncertainty on the $B$ background 
components. 
Finally, the systematic errors for the direct \CP asymmetries $A_{+-}$ and 
$A_{-+}$ (see definitions in Eq.~\ref{eq:Asy1} and Eq.~\ref{eq:Asy2}) are 
$\pm{0.16}$ and $\pm{0.09}$ respectively.

\begin{table}[htb]
\caption{Summary of the systematic uncertainties on the \CP and non-\CP
observables. The individual systematic errors are added in quadrature.}

\begin{center}
\begin{tabular}{l|c|c|c|c|c|c}
\hline
\hline
{\bf Type of systematic error } 
& $\AcprhoK$ & $\Acprhopi$ & $C_{\rho\pi}$  & $\dC_{\rho\pi}$ 
& $S_{\rho\pi}$ & $\dS_{\rho\pi}$ \\

\hline
$\Delta m \pm 0.008 \ps^{-1}$&
0.000 & 0.000 & 0.005 & 0.005 & 0.001 & 0.001 \\

$\tau \pm 0.016 \ps$&
0.001 & 0.000 & 0.002 & 0.002 & 0.002 & 0.002 \\

Time resolution   &
 0.001 & 0.001 & 0.003 & 0.002 &  0.005 &  0.006 \\

Tagging&
 0.001 & 0.001 & 0.021 & 0.012 & 0.012 & 0.012 \\

Fraction of misrec. signal events $(\pm 25 \%)$ &
 0.002 & 0.002 & 0.017 & 0.004 & 0.002 & 0.004 \\

Fraction of events with  misrec. $\rho$ charge $(\pm 25 \%)$&
 0.000 & 0.002 & 0.001 & 0.004 & 0.000 & 0.000 \\

Fitting procedure (bias) &
 0.012 & 0.001 & 0.038 & 0.033 & 0.020 & 0.015 \\

NN output signal ($B^{0} \rightarrow D^{-} \rho^{+}$) &
 0.002 & 0.002 & 0.006 & 0.006 & 0.009 & 0.013\\

NN output continuum (off-resonance data) &
 0.002 & 0.002  & 0.001 & 0.001 & 0.001 & 0.001 \\	

PDF for $\mes$ and $\de$ &
 0.006 & 0.003 & 0.009 & 0.003 & 0.005 & 0.017 \\

Particle identification&
 0.010 & 0.010 & 0.008 & 0.014 & 0.013 & 0.001 \\

$B$-backgrounds &
 0.109 & 0.065 & 0.077 & 0.107 & 0.063 & 0.035 \\
\hline
Total  & 
0.110 & 0.065 & 0.091 & 0.114 & 0.069  & 0.046\\
\hline
\hline
\end{tabular}
\end{center}
\label{tab:sys_table}
\end{table}

%% file: Summary.tex
\section{Summary}
\label{sec:Summary}
 
With a data sample of 88 million $B\Bbar$ pairs, collected
between January 2000 and June 2002 by the \babar\ detector at the \pep2 asymmetric-energy 
$B$~Factory at SLAC,  we find $413^{+34}_{-33}$~(stat)~$\rhopi$ 
and $147^{+22}_{-21}$~(stat) $\rho K$ events and  we obtain the following 
preliminary measurements of the \CP~violation parameters:
\begin{eqnarray*}
 \AcprhoK = 0.19\,\pm 0.14\,\,\,{\rm(stat)}\,\,\, \pm{0.11}\,\,\, {\rm(syst)},
& & 
\Acprhopi = -0.22\,\,\,\pm 0.08\,\,\,{\rm(stat)}\,\,\, \pm{0.07}\,\,\, {\rm(syst)}, \\ 
 C_{\rho\pi} \,\,\, = \,\,\,0.45\,\,\,^{+0.18}_{-0.19}\,\,\,{\rm(stat)}\,\,\, \pm{0.09}\,\,\, {\rm(syst)}, 
& & 
 S_{\rho\pi}\,\, =\,\,\,\,\, 0.16\,\,\,\pm 0.25\,\,\,{\rm(stat)}\,\,\, \pm{0.07}\,\,\, {\rm(syst)},
\end{eqnarray*}
and of  the other parameters in the description of the $\Bz(\Bzb) \to \rhopi$
decay-time dependence:
\begin{eqnarray*}
\dC_{\rho\pi} \, = \, 0.38\,\,\,^{+0.19}_{-0.20}\,\,\,{\rm(stat)}\,\,\, \pm{0.11}\,\,\, {\rm(syst)},
&&
\dS_{\rho\pi} =0.15\,\,\,\pm 0.26\,\,\,{\rm(stat)}\,\,\, \pm{0.05}\,\,\, {\rm(syst)}.
\end{eqnarray*}
For the asymmetries $A_{+-}$ and $A_{-+}$, which probe direct \CP~violation, we measure
\begin{eqnarray*}
A_{+-} \, = \, -0.82\pm 0.31\,\,\,{\rm(stat)}\,\,\, \pm{0.16}\,\,\, {\rm(syst)}, &&
A_{-+} \, = \, -0.11\pm 0.16 \,\,\,{\rm(stat)}\,\,\, \pm{0.09}\,\,\, {\rm(syst)}.
\end{eqnarray*}

%% file: pubboard/acknowledgements.tex
We are grateful for the 
extraordinary contributions of our \pep2\ colleagues in
achieving the excellent luminosity and machine conditions
that have made this work possible.
The success of this project also relies critically on the 
expertise and dedication of the computing organizations that 
support \babar.
The collaborating institutions wish to thank 
SLAC for its support and the kind hospitality extended to them. 
This work is supported by the
US Department of Energy
and National Science Foundation, the
Natural Sciences and Engineering Research Council (Canada),
Institute of High Energy Physics (China), the
Commissariat \`a l'Energie Atomique and
Institut National de Physique Nucl\'eaire et de Physique des Particules
(France), the
Bundesministerium f\"ur Bildung und Forschung and
Deutsche Forschungsgemeinschaft
(Germany), the
Istituto Nazionale di Fisica Nucleare (Italy),
the Research Council of Norway, the
Ministry of Science and Technology of the Russian Federation, and the
Particle Physics and Astronomy Research Council (United Kingdom). 
Individuals have received support from 
the A. P. Sloan Foundation, 
the Research Corporation,
and the Alexander von Humboldt Foundation.